# A Monte Carlo Simulation Study of the Mechanical and Conformational Properties of Networks of Helical Polymers. I General Concepts

*Gustavo A. Carri*[*], *Richard Batman, Vikas Varshney and Taner E. Dirama*

Department of Polymer Science and The Maurice Morton Institute of Polymer Science, The University of Akron, Akron, OH 44325-3909

## ABSTRACT

We study the mechanical and conformational properties of networks of helical polymers with a combination of Monte Carlo simulations based on the Wang-Landau algorithm and the Three-chain Model. We find that the stress-strain behavior of these networks has novel features not observed in typical networks made of synthetic polymers. In particular, we find that as these networks are stretched they first strengthen, then soften and finally strengthen again. This non-monotonic behavior of the stress correlates with the one of the helical content and is rationalized by the *elongation-induced* formation and melting of the helical structure of the polymer. We complement these results with a study of other conformational properties of the polymer strands that clarify the molecular mechanisms behind the mechanical behavior of these networks. Finally, we present a qualitative comparison of some of our results with the theoretical ones recently reported by Kutter and Terentjev.

[*] To whom any correspondence should be addressed. Electronic mail: gac@uakron.edu Telephone: (330) 972-7509. Fax: (330) 972-5290.

1. INTRODUCTION

The macromolecular character of polymers together with their ability to change their conformations upon the application of an external mechanical force makes their elastic behavior unique. Indeed, elasticity is one of the most important properties of polymeric materials and has been extensively studied since the ground-breaking experiments on network thermo-elasticity by Gough and Joule.[1-4] Networks made of polymers, called elastomers, are known for their large deformability with almost complete recoverability, behavior not observed in other materials like thermoplastics, metals or ceramics. This unique kind of elasticity is a consequence of three molecular characteristics of elastomers: the macromolecular nature of polymer molecules, their ability to alter their conformations and the formation of a network structure via cross links.[5]

The study of polymer networks has evolved through three different and complementary channels: experimental, theoretical and computer simulation studies. Doubtlessly, the most important of the three is the experimental one which has been the primary method used to study elastomers and has led our understanding of polymer networks to its current level. Among the different experimental methods available, mechanical testing of elastomers under uniaxial extension and compression are, perhaps, the most commonly used methods and have provided a substantial amount of information about these systems. However, they are not the only loading conditions available. Indeed, biaxial extension, shear and torsion are other possible loading conditions that lead to further and deeper information about the material. Other characterization techniques include swelling experiments, optical and spectroscopic methods like infrared dichroism, fluorescence polarization and polarized infrared spectroscopy, microscopy like scanning tunneling and atomic force microscopy, Nuclear Magnetic Resonance, small-angle neutron and Brillouin scattering, and other methods. A very comprehensive review of these experimental techniques as applied to elastomers was recently compiled by Mark.[6] From a theoretical perspective, many theories have been developed that attempt to connect the molecular properties of the material like intermolecular effects and, crosslink density and functionality with the macroscopic behavior like the stress-strain relationship. The first theories were developed by Kuhn,[7] Treolar,[2,8] James and Guth,[9] Wall[10] and Flory.[11] Further extensions of these theories, such as the Constraint Junction Model originally proposed by Ronca and Allegra,[12] and the Slip-link Model developed by Graessley,[13] Ball et al.[14] and, Edwards and Vilgis[15] accounted for intermolecular effects. A review of these first models has been recently compiled by Erman and Mark.[16] More recently, substantial progress has been made by Vilgis,[17]

Heinrich,[18] Rubinstein,[19] Terentjev[20] and Schweizer[21] among other researchers. Finally, computer simulations have also been employed to enhance our understanding of elastomers. The most common method employed to simulate the behavior of networks is a combination of a single-chain simulation and theory.[16] Specifically, a model of a single polymer is built and simulated using Monte Carlo algorithms. The output of the simulation is the radial distribution function of the end-to-end distance which is used as input for the standard Three-chain Model of rubberlike elasticity.[2,8] This model provides the stress-strain behavior of the network. This approach has been applied to many polymers like Polyethylene, Poly(dimethylsiloxane) and Poly(oxymethylene) among others.[16] Another, more recent, approach has been the use of Molecular Dynamics (MD) simulations to study the behavior of elastomers. In particular, Grest and Kremer have done extensive studies on networks using MD simulations.[22]

The aforementioned list of theoretical, experimental and computer simulation studies is by no means exhaustive. However, it might lead the reader to believe that the field of rubberlike elasticity is almost fully understood; conclusion that is far from the truth. Indeed, there are many systems of current scientific interest that are not understood at present; for example networks having multimodal distributions of network chain lengths, reversibly cross-linked materials and elastomers crosslinked in solution, or new ways to reinforcement like exfoliated clays and rubbery particles for toughening ceramics.[5] Biologically inspired ("biomimetic") elastomers and bioelastomers are not understood either. In particular, there are very few studies on protein bioelastomers like elastin and resilin when compared to the abundance of studies on synthetic elastomers.[16,23,24] However, the elasticity of *single* biopolymers has attracted a lot of attention in recent years thanks to the development of a series of experimental methods referred to as Single Molecule Force Spectroscopy (SMFS). These methods have provided a novel perspective on the structure of macromolecules and the determinants of their mechanical stability. Some of the macromolecules studied using SMFS include biological molecules like RNA[25] and DNA,[26] polysaccharides like dextran[27] and xanthan,[28] the muscle protein Titin,[29] the extracellular matrix protein Tenascin[30] and others, as well as synthetic polymers like Polyethylene glycol,[31] Poly(vinyl alcohol)[32] and Poly L-glutamic acid.[33]

The use of biomimetic or biological polymers to build novel materials is already underway.[34] In the particular case of networks these more complex polymers will clearly add a new dimension to the physics of elastomers by bringing elements characteristic of biological molecules to rubberlike elasticity. For example, biopolymers like polypeptides and proteins are known to form secondary structures like α-helices, β-sheets and β-barrels, and tertiary structures. The presence of these new elements in the chemical structure of the polymer strands together with the three molecular characteristics of elastomers mentioned in the first paragraph will clearly enrich the physics of these systems. Indeed, upon the application of an external mechanical force the response of the network will be more complex than the one of a typical synthetic elastomer due to the internal (secondary and tertiary) structure of the polymers. For example, the application of a force will not only reduce the entropy of the polymer strands but also, might stabilize or destabilize the secondary and tertiary structures which, in turn, will add to the response of the network to the applied force. This will clearly affect the stress-strain relationship of the network. In addition, the thermo-elastic behavior will be enriched since an increase in temperature will "melt or soften" the secondary and tertiary structures of the biopolymer thus, making the network *softer*. Similarly, solvent is expected to have interesting effects on the mechanical properties of the network.

The myriad of possibilities offered by the coupling of biological structures with elastomeric materials motivated us to study these systems from a modeling perspective. However, biopolymers can be structurally very complex. Therefore, in this first article we focus on biopolymers with the simplest possible secondary structure, α-helices, and explore the *macroscopic* consequences. Specifically, we focus on biopolymers called homopolypeptides which undergo the helix-coil transition with changes in temperature or solvent quality. This transition has been extensively studied in the literature.[35,36] During the transition, the polymer adopts an α-helical conformation at low temperatures while, at high temperatures, the polymer is a random coil. This suggests that at low temperatures the elastomer should behave as a network of rigid rods while, at high temperatures, it should behave as a typical network of flexible chains. However, the behavior turns out to be more complex. The transition from one behavior to the other one occurs at temperatures close to the helix-coil transition temperature. Kutter and Terenjtev have recently published theoretical results for networks of helicogenic molecules.[37] We will compare our computer simulation results to their theoretical ones in this article.

We carry out our study with a combination of single-chain simulations and the Three-chain Model. The single-chain simulations provide the radial distribution function of the end-to-end distance which we use as input for the standard Three-chain Model. The model for the single-chain (homopolypeptide) simulation was recently developed by us[36] and will be summarized in the next section. The Monte Carlo algorithm employed to solve the model is the one developed by Wang and Landau.[38] With these results and the Three-chain Model we compute the stress-strain relationship of the network and other conformational properties of the polymer strands at different temperatures.

This article is organized as follows. In the next section, we describe our simulation protocol. First, we provide a comprehensive description of the model used to describe helicogenic polymers. Afterward, we describe the simulation methodology which is based on Monte Carlo simulations using the Wang-Landau algorithm. We also review the Three-chain Model briefly. In the next section we present our results for the stress-strain curves at three different temperatures and rationalize the effect of temperature and strain on different equilibrium properties. Moreover, for the purpose of making our study more balanced and objective, we present a *qualitative* comparison of our simulation results with the theoretical ones of Kutter and Terentjev. Finally, we conclude the present article by summarizing the most important findings of our work and with the appropriate acknowledgements.

2. SIMULATION PROTOCOL AND THEORETICAL MODEL

2.1. THE MODEL

We describe a helicogenic polymer with the Freely Rotating Chain model where each bead represents an amino acid residue. The intramolecular interactions between pairs of beads are modeled with a hardcore potential energy and the tendency towards the helical conformation is modeled using a criterion based on the concept of torsion of a curve as described below.

Helicogenic polymers are known to undergo the helix-coil transition upon a change in temperature or solvent quality. One of the most important characteristics of this transition is its *cooperativity*[35] which emerges from the formation of a hydrogen bond between the pair of residues $i$ and $i+4$. This, in turn, constrains the spatial positions and orientations of residues $i+1$, $i+2$ and $i+3$. To capture this cooperative nature of the helix-coil transition we proposed a criterion based on the concept of "torsion" of a curve which is a well defined mathematical quantity. Explicitly, this concept is employed as a criterion to determine the conformational state (helix or coil) of each bead. The torsion of a curve parameterized by the vectorial field $\mathbf{r}(x)$, which could be visualized as a continuous representation of the polymer chain, is defined as follows

$$\chi(x) = \frac{(\mathbf{r}'(x), \mathbf{r}''(x), \mathbf{r}'''(x))}{\left| [\mathbf{r}'(x), \mathbf{r}''(x)] \right|^2}, \tag{1}$$

where $x$ is the arc of length parameter that can take any value in the interval $[0, L]$, $L$ being the total contour length of the chain, $\mathbf{r}'(x)$, $\mathbf{r}''(x)$ and $\mathbf{r}'''(x)$ are the first, second and third order derivatives of $\mathbf{r}(x)$, respectively. The square brackets and parenthesis indicate vectorial and scalar triple product (i.e. $(\mathbf{A}, \mathbf{B}, \mathbf{C}) = \mathbf{A} \bullet (\mathbf{B} \times \mathbf{C})$), respectively. This definition of torsion is also valid for the discrete representation of the polymer chain used in the simulation; the only difference is that the derivatives of the field must be approximated using finite differences. For example, the first order derivative of the field on the $i$-th bead is

$$\mathbf{r}'(i) \approx \frac{\mathbf{r}(i+1) - \mathbf{r}(i-1)}{2l_K}, \tag{2}$$

where $l_K$ is the bond (Kuhn) length and $\mathbf{r}(i)$ is the position of the i-th bead. Similar expressions are also available for the other derivatives.[39]

It is well known that a helical curve has a fixed torsion.[40] If we try to superimpose the beads that represent the polymer chain onto this imaginary helical curve, these beads should also have a constant value of torsion. We call this value *the torsion of the perfect helix* $\chi_{Helix}$ and was chosen to be 0.87 which corresponds to two consecutive dihedral angles of +90 degrees. Consequently, we defined the following criterion: "a bead has a helical conformation if the value of its torsion differs from the torsion of the perfect helix, $\chi_{Helix}$, by less than a certain cutoff value, $\chi_{Cutoff}$". The cutoff $\chi_{Cutoff}$ was set to 0.001. This value ensures that there is only *one* configuration of the chain that corresponds to the "helical state". If this criterion is satisfied, then the helical bead carries a negative enthalpy, called $C$, which stabilizes the helical conformation; otherwise, the bead is in the random coil state which is the reference state of the system. The enthalpic parameter $C$ is related to the standard parameter *s* of helix-coil transition theory[35] as follows

$$s = \exp(-\Delta A / k_B T) \quad \text{where} \quad \Delta A = C\text{-}T\Delta S. \tag{3}$$

It provides the enthalpic contribution that arises from the formation of a hydrogen bond and is assumed to be constant in our model. Furthermore, $\Delta S$ is the decrease in the entropy of the residue incorporated into a helical sequence due to the formation of a new hydrogen bond. The origin of ΔS in our model arises from the freely rotating chain (FRC) model and the constraints in the dihedral angles. Both parameters *C* and *ΔS* are negative in eq 3. We chose the value of *C* to be -1300 K so that the helix-coil transition temperature is close to 300K.

2.2. SIMULATION METHODOLOGY

We used the Freely Rotating Chain model where the bond length was set equal to 1.53 (in arbitrary units) and the bond angle was 109.3 degrees. Both parameters were kept constant during the simulation. The concept of torsion, as explained in the previous subsection, was used to determine the conformation of each bead for each configuration of the chain. The torsion of the perfect helix was chosen as 0.87 which corresponds to two consecutive dihedral angles of +90 degrees and the chosen cutoff value was 0.001. The values for the bond length; bond angle and perfect torsion were chosen arbitrarily for convenience and not with the purpose of mimicking the properties of any known polymer in particular.

The initial configuration of the polymer chain was generated randomly. The first bead was placed at origin and did not move during the simulation. The initial position of the second bead was along X-axis and at a distance equal to 1.53 from the origin. The third bead was located on the X-Y plane so that the bond angle was 109.3 degrees and, moreover, the second bond had a positive projection onto the Y-axis. The positions of all other beads were computed using random dihedral angles with respect to the previous three beads. These angles were taken from a prescribed finite set of sixty four possible values, $\phi = 90 + (m\pi/32)$ where $\phi$ represents the dihedral angle and *m* varies from *0* to *63*. This was done for the purpose of having a finite estimate for the density of states as explained below.

Pivot moves were used to change the configuration of the chain. In the pivot move, the *i*-th bead is selected randomly and the rest of the polymer (beads *i*+1 to *n)* is rotated around the bond between beads *i-1* and *i* by a randomly chosen angle.

The Monte Carlo algorithm chosen was the one developed by Wang and Landau.[38] This algorithm generates a random walk in energy space with a probability proportional to the reciprocal of the density of states, *g(E)*, and leads to a flat energy histogram. This method has the advantage of escaping local energy minima and exploring the free energy landscape efficiently which, in turn, leads to an accurate estimate of the density of states, *g(E)*, for any system of interest. Once the density of states is known, all the statistical properties can be evaluated using standard formulae from Statistical Mechanics.

For the purposes of this article, we employ the Wang-Landau algorithm to explore energy (arising from *the number of beads in the helical conformation*) space which is required for the estimation of the density of states. The energy can adopt discrete values ranging from 0 to $(n-4)C$ in steps of $C$ where $n$ is the number of beads. So, in our case, the density of states is a function of only one variable: the energy $E$ due to the beads in the helical state.

We now review the Wang-Landau[38] algorithm. At the beginning of the simulation, the density of states $g(E)$, which is unknown *a priori*, is initialized to one for all possible values of the energy. The random walk is then started by changing the configuration of the polymer. The transition probability for switching the polymer configuration from $\{E_i\}$ to $\{E_f\}$ is

$$\text{Prob}(E_i \to E_f) = \min\left(1, \frac{g(E_i)}{g(E_f)}\right). \tag{4}$$

Each time a move is accepted, the density of states of the new configuration is updated by multiplying the existing value by a modification factor $f$, i.e., $g(E) \to g(E) \times f$. However, if the move is rejected, then the density of states of the old configuration is updated. This modification allows the random walk to explore energy space quickly and efficiently. The starting value of $f$ was taken to be $e^1 (=2.71828)$. After a move is completed, the corresponding histogram $H(E)$ is updated along with the modification of the density of states. Once the histogram is "flat" within some tolerance, the value of $f$ is modified as follows $f_{new} = \sqrt{f_{old}}$. At this point, the histogram is reset to zero and the above procedure is started again with the updated modification factor. This procedure is repeated until the value of $f$ is very close to 1. We stopped our simulations when $f - 1$ became smaller than $10^{-7}$.

Using the density of states various quantities can be calculated with formulas from Statistical Mechanics. For example, the canonical partition function and Helmholtz free energy are

$$Z(T) = \sum_E g(E) e^{-\beta E}, \tag{5}$$

$$F(T) = -T \ln\left(\sum_E g(E) e^{-\beta E}\right), \tag{6}$$

where $F(T)$ is in units of Boltzmann's constant, $k_B$, and $\beta=T^{-1}$. Apart from these thermodynamic quantities, the ensemble average of any other quantity of interest can also be calculated using following equation

$$\langle A(T) \rangle = \frac{\sum_E A(E) g(E) e^{-\beta E}}{\sum_E g(E) e^{-\beta E}}. \qquad (7)$$

In our case, we are interested in the radial distribution function, the helical content which is defined as the fraction of the polymer in the helical conformation and, the average length and number of helical strands. The computation of these quantities was done as follows. The first step was to compute the density of states using the Wang-Landau sampling scheme described before. Once the density of states was known, we generated $10^9$ configurations of the chain and stored the values of the average end-to-end distance squared, helical content and other quantities for each configuration. Since the end-to-end distance must be between zero and the maximum end-to-end distance, we divided this maximum distance in 190 bins. So that when the data of each of the $10^9$ configurations were read, we used the end-to-end distance to locate the bin to which the configuration belonged. Then, the value of the property of interest in that particular bin was updated. For example, in the particular case of the radial distribution function we increased the value stored in the bin by the density of states corresponding to that bin times the Boltzmann weight at the temperature of interest. The final result was then normalized. Similar procedures were used for the other conformational properties.

### 2.3. THE THREE-CHAIN MODEL

The three-chain model assumes that interchain interactions are independent of deformation and averages the free energies of chains in three orthogonal orientations. The model considers three effective chains with end-to-end distances $R_i$ ($i$=x,y,z) parallel to the coordinate axes that are deformed in the affine limit at constant volume. The macroscopic deformations of the network are defined as elongations $\lambda_i = \frac{L_i}{L_{io}}$ where $L_i$ and $L_{io}$ indicate the deformed and undeformed dimensions of the network in the $i$-th direction, respectively. For uniaxial extension the conservation of volume implies that $\lambda_x = \lambda, \lambda_y = \lambda_z = \lambda^{-\frac{1}{2}}$. Thus the total free energy of the network made of $\phi$ chains per unit volume is given by the equation

$$\Delta f_{net} = \phi \left( \frac{f(R_o \lambda)}{3} + \frac{2 f\left(R_o \lambda^{-\frac{1}{2}}\right)}{3} - f(R_o) \right) \tag{8}$$

$R_o$ is the average chain dimensions of the network chains in the undeformed state and $f(x)$ is the single-chain free energy (in units of Boltzmann constant) given by $f(x) = -T \ln(W(x))$ where W(x) is the *probability distribution* of the end-to-end distance obtained from the Monte Carlo simulation. The derivative of $\Delta f_{net}$ with respect to $\lambda$ gives the stress-strain relationship.

3. RESULTS AND DISCUSSION

Let us start by analyzing the radial distribution function (rdf) obtained from our Monte Carlo simulation study of a single chain. Figure 1 shows the rdf as a function of the end-to-end distance for a chain with 30 beads and for three different temperatures: 250, 300 and 350 K. The helix-coil transition temperature for a 30 bead chain is 311 K and is mainly determined by the parameter $C$ (=-1300 K) of the model. Therefore, the three temperatures chosen cover all the range of possible behaviors of the chain from the helical structure at low temperatures to the random coil conformation at high temperatures. Figure 1 shows that as the temperature is reduced, the peak in the rdf shifts towards larger values of the end-to-end distance. This behavior is the expected one since a decrease in temperature favors the formation of helical sequences thus, making the chain stiffer. However, it is interesting to observe the presence of two peaks in the rdf at low temperatures. This result agrees qualitatively with previous studies by Curro and Mark on Poly(oxymethylene) (POM) with low number of skeletal bonds (20 and 40).[41] POM is known to adopt the helical conformation under some conditions.

Figure 2 shows the Helmholtz free energy obtained from the single chain Monte Carlo simulation as a function of the end-to-end distance for the three temperatures mentioned before. The plot clearly shows that a decrease in temperature shifts the minimum of the single chain free energy towards larger values of the end-to-end distance. This is a consequence of the behavior of the rdf shown in Fig. 1. However, the most interesting feature of the single chain free energy is the change in its behavior for large values of the end-to-end distance. Observe that as the temperature is reduced, the dependence of the single chain free energy on the end-to-end distance becomes stronger, i.e. the slope becomes steeper. This indicates a sharp change in the elastic behavior of the polymer and has important consequences in the free energy of the network and, consequently, in the stress-strain behavior. The free energy of the network as a function of the extension ratio, $\lambda$, is shown in Fig. 3 for the temperatures mentioned before. The sharp change in the elastic behavior of the polymer for large end-to-end distances appears as a sharp increase in the free energy of the network as a function of $\lambda$, e.g. at 300 K the increase occurs for values of $\lambda$ slightly lower than 1.5. These sharp changes in the free energy of the network point to a substantial change in the elastic behavior of the polymer strands as we discuss below.

Figures 4 and 5 show the stress and helical content, i.e. the fraction of the polymers adopting the $\alpha$-helical conformation, of the network as a function of the strain at the three aforementioned temperatures for the cases of uniaxial extension and compression. We start by analyzing the case of simple extension ($\lambda>1$). For all the cases studied the stress increases with increasing strain, as expected. However, the physical origins of this increase are more complex than in the case of synthetic polymers. Indeed, there are two kinds of forces resisting the deformation of the network: the decrease in the entropy of the system which is also present in synthetic elastomers and can be described using classical theories of rubber elasticity, and the formation of helical strands which is known to be facilitated by the application of external mechanical forces.[36] Figure 4 shows that for temperatures below the helix-coil transition temperature, 250 and 300 K in our case, the stress first increases indicating that the network gets stronger as we stretch it, then decreases which points to a softening of the network before finally increasing to very large values due to the finite extensibility of the polymer strands. At temperatures above the helix-coil transition temperature the softening of the network is not observed, i.e. the stress always increases with increasing $\lambda$. However a clear "shoulder" is observed before the final increase of the stress. Thus, this clearly indicates that the origin of the decrease in the stress for temperatures below the helix-coil transition temperature is directly related to the presence of helical strands *before* the network is stretched.

Figure 5 shows the helical content as a function of the strain. At 250 K the helical content is close to 0.9 indicating that 90 percent of the polymer strands are in the $\alpha$-helical conformation. Observe that this value remains constant until it starts to decrease for values of $\lambda$ close to 1.15. This implies that extension ratios larger than 1.15 interfere with the formation of helical strands, i.e. the elongation of the network is too large for the formation of helical strands or, in other words, the polymer strands are *overstretched* with respect to the end-to-end distance of the helix. Therefore, an increasing number of segments adopt the random coil conformation to satisfy the constraint imposed by the deformation. Consequently, the helical content decreases. It is important to note that this behavior of the helical content predicted by our simulation study was also predicted by the theoretical calculations of Kutter and Terentjev (Fig. 10 in Ref. 37). But, they used the Gaussian distribution for the description of flexible coils which did not allow us to carry out a *quantitative* comparison between theory and simulations. However, the results agree on a *qualitative* level. At 300 and 350 K the behavior of the helical content is slightly different. Indeed, upon the extension of the network the helical content first increases and then decreases in *qualitative* agreement with the results of Kutter and Terentjev (Fig. 9 in Ref. 37). The initial increase of the helical content indicates that the application of an external mechanical force first *stabilizes* the helical conformation. This can be achieved in two ways: first, the helical strands could be longer, i.e. more beads per strand, and, second, more helical strands could be formed. A further increase of the extension ratio interferes with the formation of helices and the helical content decreases. It is important to notice that the curves for the helical content do not reach zero for large values of $\lambda$. This is a limitation imposed by the number of bins (=190) used to compute the rdf, Fig. 1. A larger number of bins would solve this problem. However $10^9$ configurations would not be enough to get accurate values for the free energy and, thus, the fluctuations in the strain-stress relationship would be very large.

Using the results obtained for the helical content and the stress we can now explain the physical origins of the stress-strain behavior. For this purpose we combined the data for stress and helical content at 300 K in one plot, Fig. 6. Observe that for values of $\lambda$ slightly above 1 both the stress and the helical content increase. This implies that more and/or longer helical strands are formed as we stretch the network. We discuss this point below. The formation of new helical strands and/or the increase of the length of the helical strands decrease the entropy of the system because it removes many rotational degrees of freedom from the polymers. Thus, the decrease of the entropy of the network is faster than in the typical case of synthetic polymers (i.e. without secondary structures) and, moreover, has two contributions: first, the loss of entropy due to the elongation of the random coil segments and, second, the loss of *random coil* segments due to the stabilization of the helical conformation by the external mechanical force.

Figure 6 also shows that when the helical content is about to reach the maximum, the stress increases abruptly. This implies that for this particular value of the extension ratio the polymer strands are almost completely aligned parallel to the stretching force and all the random coil segments are almost fully stretched. Consequently, the external force is resisted by the helical strands. Since these strands are more stable due to the molecular driving forces that stabilize the helical conformation (e.g. hydrogen bonds), more force is required to overcome this thermodynamic barrier and *melt* the helices. Consequently, the stress increases. Upon further increase of $\lambda$ the network overcomes the thermodynamic barrier and, fewer and shorter helices remain. Consequently, the number of segments in the random coil conformation increases which, in turn, increases the entropy of the network. Thus, the network softens and the stress decreases. Finally, for very large deformations the polymer strands are fully stretched and the stress diverges to infinity.

The previous rationalization of the stress-strain behavior at 300 K is also valid for 350K. However, the behavior at 250 K is slightly different. Indeed, the helical content shown in Fig. 5 remains approximately constant while the stress increases. This implies that even for small values of $\lambda$ the external force is resisted by the helical strands. This is to be expected because 90 percent of the beads are in the helical conformation.

In order to understand the behavior of the helical content we plot the average length of a helical strand, $\nu$, and the average number of helical strands, $\mu$, as a function of $\lambda$ in Figs. 7 and 8. Observe that at 250 K both $\nu$ and $\mu$ are approximately constant until they decrease sharply for values of $\lambda$ larger than 1.15. Thus, the behavior of the helical content follows the ones of $\nu$ and $\mu$. These results indicate that the mechanism leading to the decrease of the helical content with increasing values of $\lambda$ involves helical strands breaking into shorter ones and *not* unwinding from the ends which is another mechanism found in the helix-coil transition of short homopolypeptides. At 300 K the situation is different. $\mu$ decreases with increasing $\lambda$ which implies that the deformation decreases the average number of helical strands. However, Fig. 7 shows an increase in $\nu$. Eventually, $\nu$ overrides the decrease of $\mu$ and the helical content increases. These results suggest that the deformation of the network tends to merge helical strands into longer and thermodynamically more stable helical strands. Again, for large deformations the helical strands do not unwind from the ends but, break into shorter helical strands. Finally, the situation changes again at 350 K. In this case both $\nu$ and $\mu$ increase with increasing λ. This leads to a substantial increase in the helical content by a factor of 5, approximately (Fig. 5) and indicate that the deformation of the network stabilizes helical strands of any length.

Let us now rationalize the results obtained for the case of compression (λ<1) briefly. In this case, the network is compressed in *one* direction. Consequently, it is stretched in the two orthogonal directions because the volume must be conserved in the Three-chain Model. Thus, all the results reported for the case of uniaxial extension like the increase in the helical content and ν with increasing λ, etc. should also appear in compression for *decreasing* values of λ. This is what all the plots show. The only difference is that the stress is now negative indicating that the system is under compression. However, all the other features observed for the case of uniaxial extension are present and can be rationalized using similar arguments to the ones employed to understand the effect of uniaxial elongation.

4. CONCLUSIONS

In this article we have studied the effect of temperature and strain on the mechanical and conformational properties of an elastomer made of helicogenic polymers. For this purpose we employed a combination of single-chain Monte Carlo simulations based on the Wang-Landau sampling scheme and the Three-chain Model of rubberlike elasticity. The helicogenic polymer was modeled with a modified Freely Rotating Chain model recently developed by us.

At the level of a single chain we found that a decrease in temperature increases the end-to-end distance. This behavior was rationalized with the argument that a decrease in temperature favors the formation of helical sequences thus, making the chain stiffer. Moreover, it was observed that at low temperature the radial distribution function displays two peaks in agreement with previous fully atomistic studies on Poly(oxymethylene). We also found that the Helmholtz free energy becomes a more sensitive function of the end-to-end distance as the temperature is reduced. This changes the elastic behavior of the polymer substantially in particular for large elongations.

For the case of the elastomer we found that the stress-strain relationship shows new features not present in the case of typical elastomers. In particular, the stress first increases indicating a strengthening of the network and then decreases before the final increase to very large values due to the finite extensibility of the polymer strands. The decrease for intermediate values of the extension (*softening* of the network) was proven to be a consequence of the *melting* of the helical structure by overstretching with respect to the end-to-end distance of the helix. The helical content was also studied. For this quantity we found that, except at very low temperatures where the helical content is approximately constant, it increases with increasing deformation. This clearly implies a stabilization of the helical conformation by the applied force. However, for large deformations of the network the helical content decreases to accommodate the constraints imposed by the applied deformation. Finally, we correlated the behaviors of the helical content and stress, and found that the sharp increase, followed by a decrease, in the stress was due to the *melting* of the helical strands by the imposed deformation. Further studies of the average number and length of the helical strands shed more light on the mechanisms behind the formation and melting of the helical strands.

Clearly, the results presented in this article will benefit from further studies. At present we only have data for three temperatures. A more detailed study that addresses a wider temperature range with more values of the temperature would be very useful since it will clarify further how the transition from low to high temperatures occurs. Moreover, it will allow us to study the thermo-elastic behavior predicted by the model.[42] Another point of interest is the effect of chain length. In this article we studied the case of 30 beads *only*. Clearly, longer chains have to be studied. In addition, the effect of solvent is also an important topic for swelling experiments. The final step would be to avoid the use of the three-chain model. This can be accomplished in a similar manner to the studies of Grest and Kremer. We have to create a model for the whole network and run the Monte Carlo simulation. All these topics will be the subjects of future studies.

5. ACKNOWLEDGMENT

This material is based upon work supported by the National Science Foundation under Grant No. CHE-0132278. Also, acknowledgement is made to The Ohio Board of Regents, Action Fund (Grant#R566).

LIST OF FIGURES

FIGURE 1: Radial distribution function as a function of the end-to-end distance for a chain with thirty beads at different temperatures. (▲) 350 K, (■) 300K and (●) 250 K.

FIGURE 2: Single chain free energy as a function of the end-to-end distance for a chain with thirty beads at different temperatures. (▲) 350 K, (■) 300K and (●) 250 K.

FIGURE 3: Network free energy as a function of $\lambda$. (▲) 350 K, (■) 300K and (●) 250 K.

FIGURE 4: Nominal stress ($\sigma^*$) as a function of $\lambda$. (▲) 350 K, (dotted line) 300K and (continuous line) 250 K.

FIGURE 5: Helical content as a function of $\lambda$. (▲) 350 K, (■) 300K and (●) 250 K.

FIGURE 6: Helical content and nominal stress ($\sigma^*$) as a function of $\lambda$ at 300K.

FIGURE 7: Average number of beads per helical strand as a function of $\lambda$. (▲) 350 K, (■) 300K and (●) 250 K.

FIGURE 8: Number of helical strands as a function of $\lambda$. (▲) 350 K, (■) 300K and (●) 250 K.

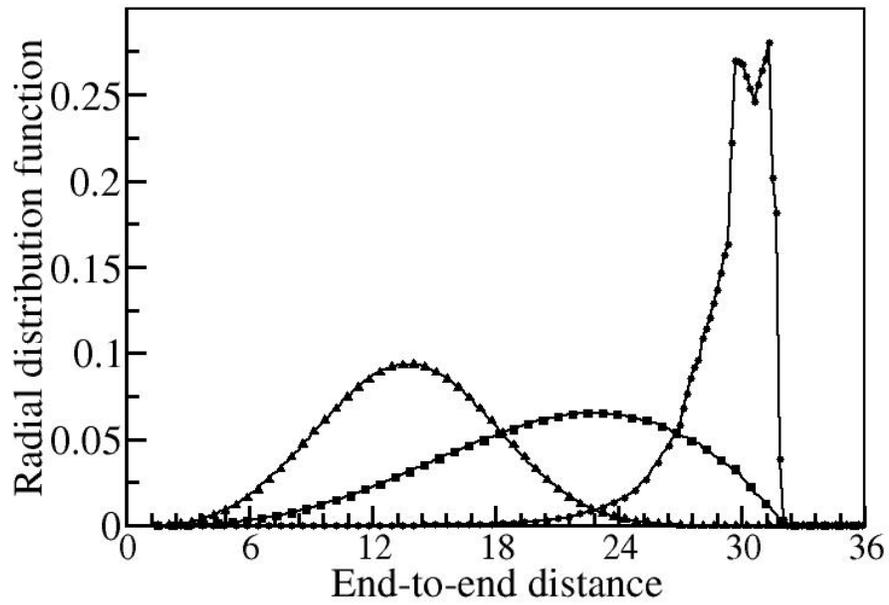

Figure 1: Radial distribution function as a function of the end-to-end distance for a chain with thirty beads at different temperatures. (▲) 350 K, (■) 300K and (●) 250 K.

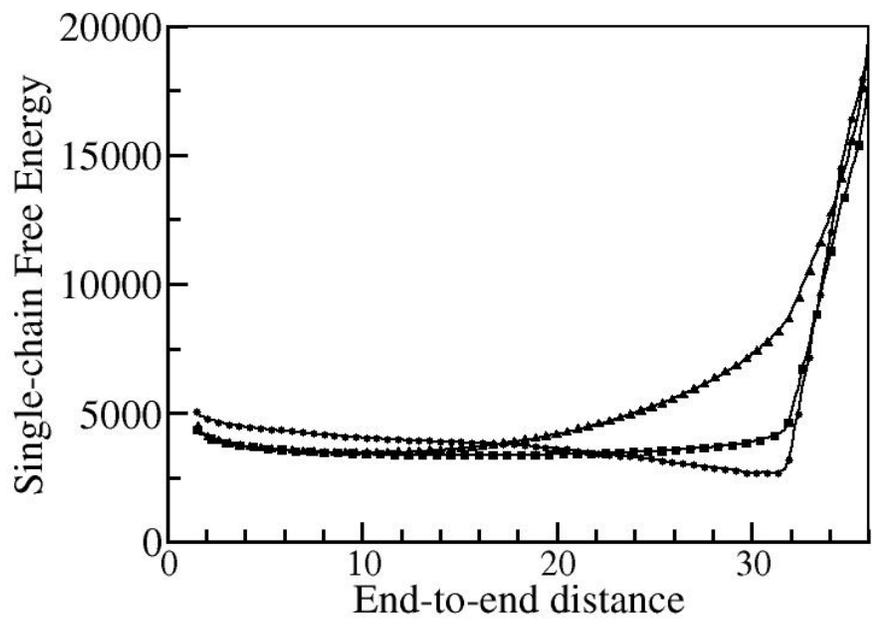

Figure 2: Single chain free energy as a function of the end-to-end distance for a chain with thirty beads at different temperatures. (▲) 350 K, (■) 300K and (●) 250 K.

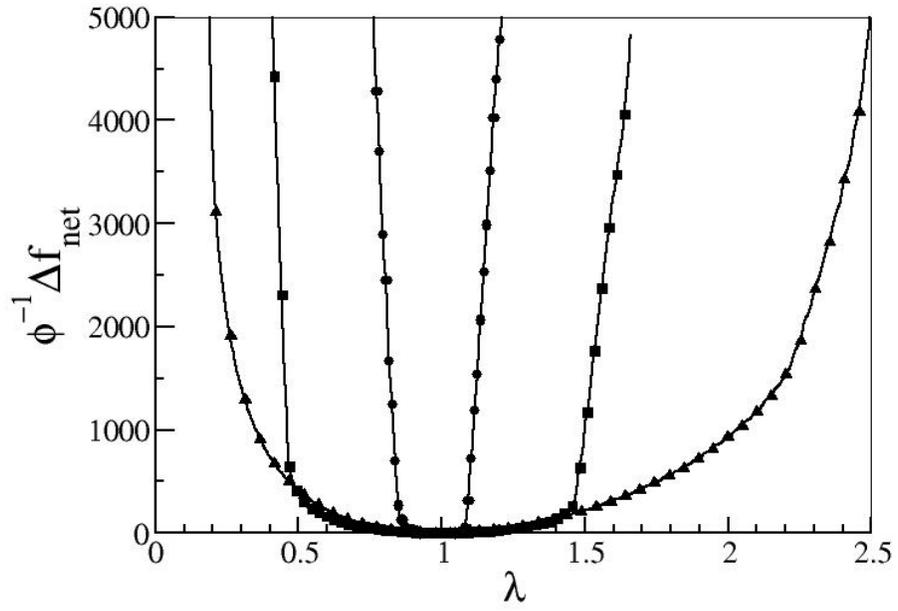

Figure 3: Network free energy as a function of λ. (▲) 350 K, (■) 300K and (●) 250 K.

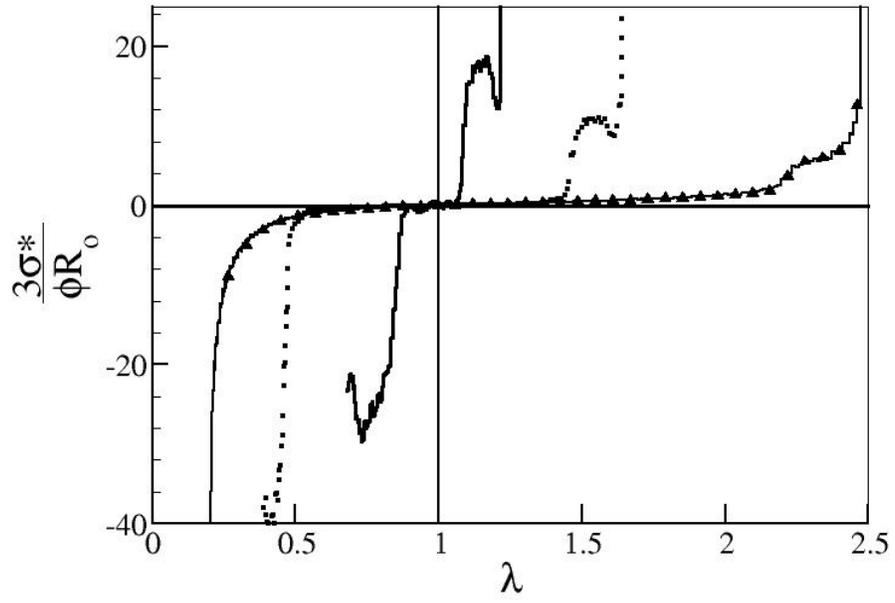

Figure 4: Nominal stress (σ*) as a function of $\lambda$. (▲) 350K, (dotted line) 300K and (continuous line) 250K.

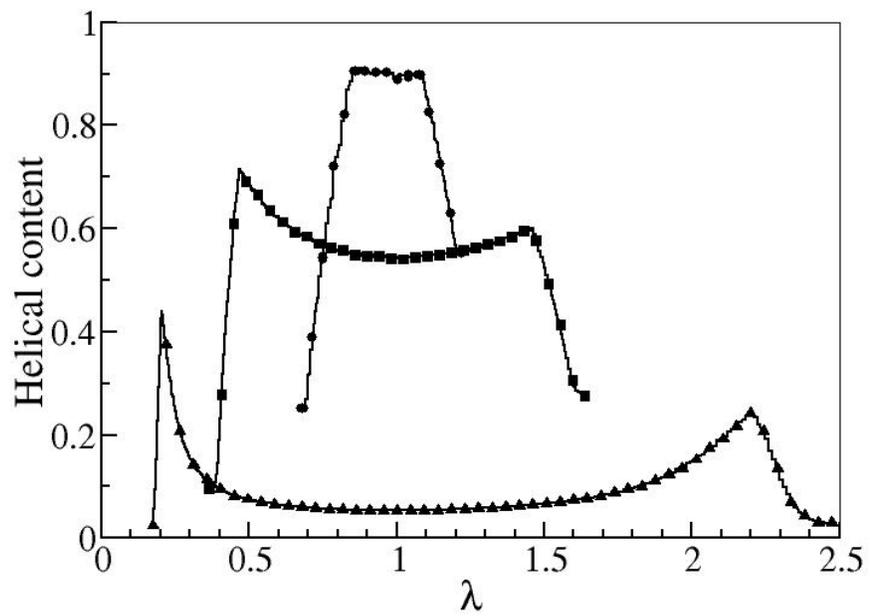

Figure 5: Helical content as a function of λ. (▲) 350 K, (■) 300K and (●) 250 K.

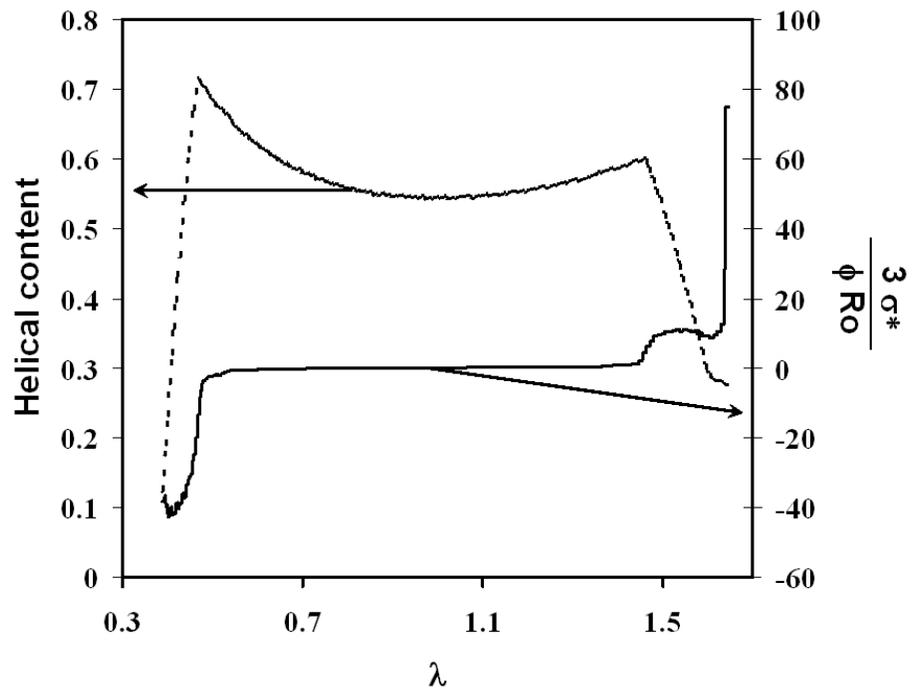

Figure 6: Helical content and nominal stress (σ*) as a function of λ at 300K.

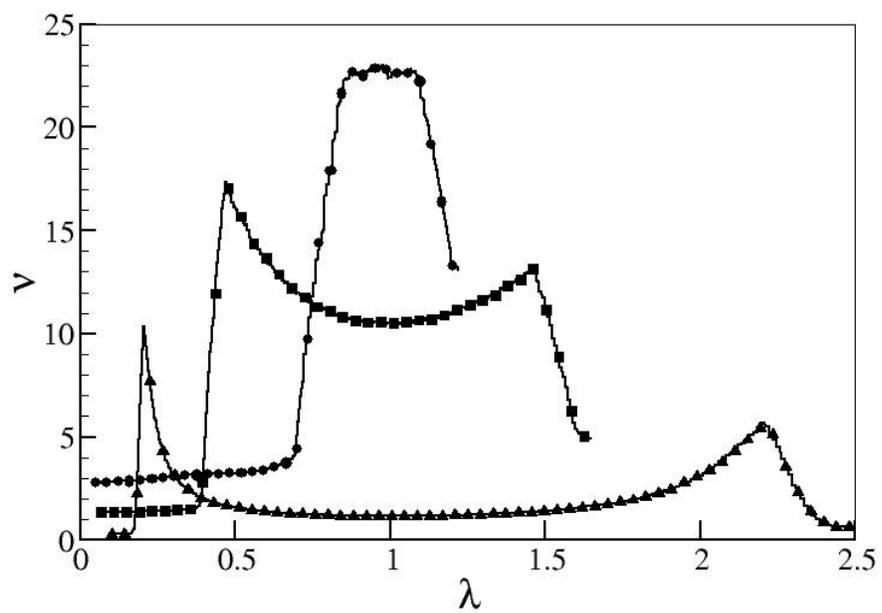

Figure 7: Average number of beads per helical strand as a function of λ. (▲) 350K, (■) 300K and (●) 250K.

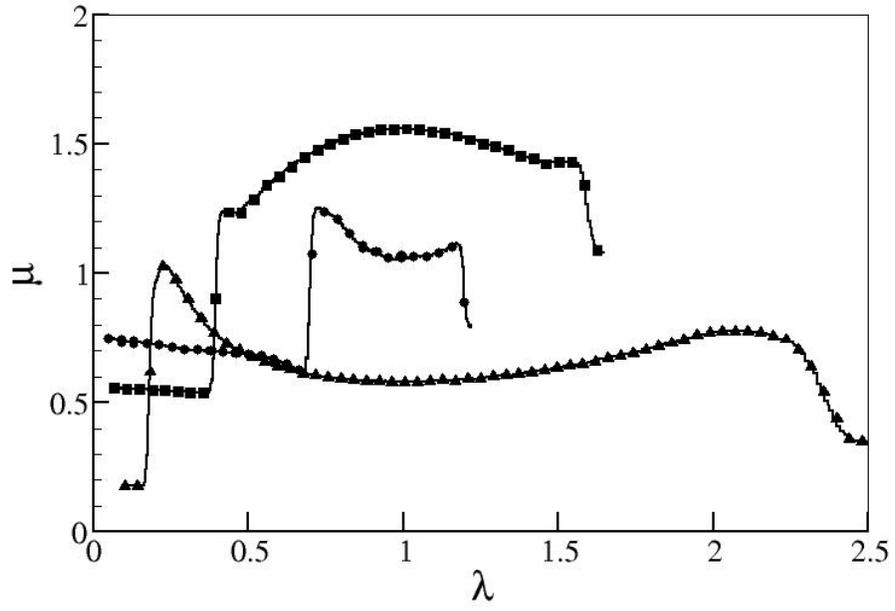

Figure 8: Number of helical strands as a function of λ. (▲) 350 K, (■) 300K and (●) 250 K.